\documentclass[reqno,11pt]{amsart}
\usepackage{amsmath, latexsym, amsfonts, amssymb, amsthm, amscd}
\usepackage{color}
\usepackage{graphics,epsf,psfrag}
\numberwithin{equation}{section}

\newtheorem{theorem}{Theorem}[section]

\newtheorem{rem}[theorem]{Remark}

\begin{document}

\title[Log-periodic critical amplitudes: a perturbative approach]{  
Log-periodic critical amplitudes: \\a perturbative approach
}

\author{Bernard Derrida}
\address{
          {   Laboratoire de Physique Statistique \\
              \'Ecole Normale Sup\'erieure,
 Universit\'e Pierre et Marie Curie, Universit\'e Paris Diderot, CNRS \\
              24, rue Lhomond,
              75231 Paris Cedex 05 - France }
}

\author{Giambattista Giacomin}
\address{
  Universit{\'e} Paris Diderot  and Laboratoire de Probabilit{\'e}s et Mod\`eles Al\'eatoires,
U.F.R.                Math\'ematiques, B\^at. Sophie Germain, 5 rue Thomas Mann, 
75205 PARIS Cedex 13 - France }

\date{\today}

\begin{abstract}
Log-periodic amplitudes appear in  the critical behavior of  a large class of systems, in particular when a discrete scale invariance is present.
Here we show how to compute  these critical amplitudes perturbatively when they
originate from a renormalization map which is close to a monomial.
In this case, the log-periodic amplitudes of  the subdominant corrections to the leading critical behavior can also be calculated.
\end{abstract}

\maketitle

\begin{center}
{\sl \normalsize Dedicated to Herbert Spohn, our colleague, friend and source of inspiration, \\
on the occasion of his $65^{\textrm{th}}$ birthday}
\end{center}

\section{Introduction}
Power laws with non integer exponents   are common in the study of scale invariant structures,   in particular  in  the theory of critical phenomena   at second order phase transitions.
In some cases, when the   scale invariance is discrete { \cite{sca1,stauffer,cas3,sca2,sca2-bis,sca3,sca4,sca5}},   the amplitude of the  powerlaw acquires a periodicity, often called log-periodic oscillations: see  { \cite{cf:Sornette,Sornette-bis}} for reviews on the topic.
These oscillatory amplitudes are usually  more difficult to  calculate {  \cite{cal0,cal1,cal2,cal3} }  than the critical exponents. When the scale invariance is associated to a renormalization group map, they can be related to some properties of the  map,
{for example to  the shape of its Julia set, see \cite{cal0,cf:CG}}.

The goal of the present paper is to show that these oscillatory amplitudes can be computed perturbatively 
 when the unperturbed map is simple enough, in our case a monomial map. 
To be more precise let us consider the following question:
take  the  map
 $x \mapsto F(x)$
with
\begin{equation}
F(x)\, =\, F_\epsilon (x)\, =\,  x^p \left(1 + \epsilon  \ G(x) \right)\, , 
\label{F-def}
\end{equation}
where
  $p=2,3, \ldots$ and the perturbation $G(x)$ { is sufficiently regular { and has} a finite limit for $x \to \infty$
  (we will actually need a condition on the decay of $G'(x)$ too, and this will be pointed out later on, see \eqref{V-def}).
If $x_*$ is the largest  real  fixed point of the map (so that\mbox{ $x_* = F(x_*)$)}, this fixed point is unstable, that is
$F'(x_*) > 1$,
 for $\epsilon>0$ sufficiently small. We will actually just focus on $F(x)$ for $x \ge x_*$.
If $F^{(n)}(x)$ denotes the $n$-th iterate of the map $F$, 
 the sequence $ \psi_n(x)$ defined by
\begin{equation}
\psi_n(x)  \, =\,  {\log F^{(n)}(x) \over p^n}\, ,
\label{psi-n-def}
\end{equation}
has a  limit
  for all $x\ge x_*$
\begin{equation}
\psi_\infty(x)  = \lim_{n \to \infty} \psi_n(x) \, .
\label{psi-inf-def}
\end{equation}
One directly sees that $\psi_\infty(x)>0$ for $x>x_*$, while $\psi_\infty(x_*)=0$.
 The existence of the limit directly follows from
\begin{equation}
\left \vert \psi_{n+1}(x) -\psi_n(x)\right \vert \, \le \, p^{-n-1} \log \left(1+ \epsilon \sup_{x\ge x_*} \vert G (x)\vert \right)\, .
\end{equation}
It is straightforward to see that the limit satisfies the functional equation
\begin{equation}
p  \, \psi_\infty(x) \, =\, {\psi_\infty(F(x))}\, ,
\label{eq-fonct}
\end{equation}
which goes under the name of  B\"ottcher equation in the iterated function literature \cite{cf:Beardon,cf:Milnor},
 and one expects  and sometime can prove -- see for example \cite{cf:BB,cf:CG} -- that 
\begin{equation}
\psi_\infty(x) \stackrel{x \searrow x_*}\sim (x-x_*)^\alpha \  B\left(-{\log(x-x_*) \over \log F'(x_*) }\right)\, ,
\label{crit-1}
\end{equation}
 where the exponent $\alpha$ is usually non-integer and $B(\cdot)$  is a strictly positive periodic function of period one.
The  exponent $\alpha $ has the  well known  expression  
\begin{equation}
\alpha \, =\, {\log p \over \log F'(x_*)}\, ,
\label{alpha}
\end{equation}
while the oscillatory amplitude  $B(y)$ is usually hard to determine.
One goal of the present paper is to show how to { compute}  this oscillatory 
amplitude perturbatively
 ({related small $\epsilon$ expansions can be found in
\cite{cf:CH}}). We will see  in Section~\ref{sec:main} that all the subdominant corrections to (\ref{crit-1}) can be calculated as well since they all can be expressed in terms of the periodic function $B(\cdot)$.
\medskip

\begin{rem}
\rm  
   \eqref{crit-1} is easily guessed: if one assumes that \mbox{ $\psi_\infty(x) \sim (x-x_*)^\alpha  C(x-x_*)$} as $x \searrow x_*$,
  with $y \mapsto C(y)$ defined for $y\in (0, y_0)$ for an arbitrary choice of $y_0>0$, continuous and such that $ 0 < C_{\rm min} < C(y) < C_{\rm max}$, one gets from  \eqref{eq-fonct} that
\begin{equation}
(x-x_*)^\alpha \  C(x- x_*) \, \sim \,  \frac {F'(x_*)^\alpha} {p} \Big( x-x_*\Big)^\alpha     C\Big(F'(x_*)( x-x_*)\Big) \, ,
\end{equation}
and this implies the value (\ref{alpha}) of $\alpha$  and the fact $C\Big(F'(x_*)y\Big)= C(y)$ for every $y$, that is that
$C(\cdot)$ is multiplicatively periodic of period $F'(x_*)$. Equivalently, 
$C(y)=B(-{\log(x-x_*) / \log F'(x_*) })$, with $B(\cdot)$ periodic of period one,  as in (\ref{crit-1}).
\end{rem}
\medskip 

\begin{rem} \rm
In what follows we 
will often assume that
 \begin{equation}
G(1)\, =\, 0\, , 
\label{G1}
\end{equation}
so that  $
x_*=  1$
 when $\epsilon$   is sufficiently small.
 There is no loss of generality in doing so because the general case  can be reduced to $G(1)=0$
by the change of variable $x \mapsto x/x_*$ and $x_*$ can easily be computed perturbatively in $\epsilon$
 \begin{equation}
\label{eq:rem1.1}
x_*\, =\,  1 - \epsilon \,  {G(1) \over p-1} + \epsilon^2  \, {p \, G(1)^2 + 2 \,  G(1)  \, G'(1) \over 2 (p-1)^2 } + ... 
\end{equation}
\end{rem}


\medskip

In this paper we  will also discuss
 the case where
the map $F(x)$ has the form
\begin{equation}
\label{large-p2}
F(x)\, =\, F_p(x)\, =\, q x^p +(1-q) P(x)\, ,
\end{equation}
where $p$ is a large positive  integer,
 $q\in(0,1)$ and $P(\cdot)$ is a polynomial such that $P(1)=1$.
 We aim at understanding what the oscillatory behavior becomes for large  {$p$ (when} $P(\cdot)$ and $q$ are 
 fixed). 
The assumptions guarantee that, for $p$ sufficiently large, $1$ is the largest fixed point (and it is unstable). 
Maps in this class are for example $(x^p+1)/2$ and
$(x^p+2x^2-x)/3$. We will see that for large $p$
  the oscillatory amplitude can be sharply estimated. In a nutshell, the reason why the map
  becomes amenable for  large $p$  is that sharp estimates of the iterated map 
$F^{(n)} (x)$ are easy
  as long as $x$ is very close to $x_*$ -- the linear regime -- and when $x$ is large, because the map
  there is close to $q x^p$. The only difficult  part of the iteration is in between
  these two regimes: we will see however that for $p$ large this intermediate regime essentially reduces to one iteration
  which one performs explicitly.  

\section{Two examples}
Let us discuss here briefly two different examples where    the above question is relevant.

\subsection{The wetting problem on a hierarchical lattice}
One of the developments in  the { theory}  of critical phenomena by the renormalization group approach was to { study}   statistical mechanical models on hierarchical lattices { \cite{hie1,hie2,hie3,hie4,hie5}}. These  lattices 
  are usually constructed by a { recursive} procedure  which is at the origin of 
 a discrete scale invariance and which allows one  to write exact renormalization transformations.

 Here we consider the so-called diamond lattice { \cite{hie2,hie4}}  (in fact a generalization of it) which is constructed as follows.
One starts with a lattice composed of a single bond between two sites $A$ and $B$.
Then at  every  iteration step  each bond of  the previous step  is replaced by  $p \times b$ bonds, namely by $b$ parallel
paths of $p$ bonds each.  Therefore if, at step $0$, one starts with a single bond between $A$ and $B$, then, after $n$ iterations, there are $(p b)^n$ bonds in the lattice, $b^{(p^n-1)/(p-1)}$ directed paths from $A$ to $B$,  and each of these directed paths  has a length $L_n=p^n$.

One of the simplest critical phenomenon one  may consider on  such a hierarchical lattice is the wetting transition { \cite{wet1,wet2,wet3}}:
in the wetting problem, all the bonds along  a single path (the special path)  from $A$ to $B$ have an energy $-a$ whereas all the  remaining  bonds on the lattice  have energy $0$. The energy $E_W$ of a path $W$  from $A$ to $B$ is simply the number of bonds it  has in common with the special path.
 Equivalently this  energy $E_W$ is the sum of the energies of the bonds visited by the path $W$. Then the partition function,
at temperature $T$, is defined as
\begin{equation}
Z(T)\, =\,  \sum_W \exp \left( - {E_W \over T} \right),
\end{equation}
{ where the sum is over all the  $ b^{p^n-1 \over p-1}$   paths $W$.}

When the temperature varies, one observes a transition between a bound phase, where typical paths have a non zero fraction of their { length} in common with the special path, to a free phase where this fraction vanishes.
At infinite temperature, all paths    have weight $1$ and the partition function counts simply the total number $Z_n(\infty)$ of paths 
\begin{equation}
Z_n(\infty)\, =\,  b^{p^n-1 \over p-1} =b  \ \left(Z_{n-1}(\infty)\right)^p\, ,
\end{equation}
while at  finite temperature,  the partition function satisfies the following recursion
\begin{equation}
Z_n(T) \, =\,  Z_{n-1}(T)^p + (b-1)  \,  Z_{n-1}(\infty)^p \, .
\end{equation}
If one introduces the ratio  $x_n= Z_n(T)/Z_n(\infty)$, it follows that 
\begin{equation}
x_{n} \,=\,  F(x_{n-1}) \ \ \ \ \ {\rm where} \ \ \  F(x) \, =\,  {x^p + b-1 \over b}\, ,
\label{HL}
\end{equation}
with $x_0= \exp({a /T})$. 
At the $n$-th step the free energy  $f_n$ per  unit length is 
\begin{equation}
f_n(x_0)\, :=\, {\log Z_n(T) \over L_n} 
\, =\, {\log F^{(n)} (x_0) \over p^n}+    {\log Z_n(\infty) \over p^n}
\, .
\end{equation}
{
which implies  for the large $n$ limit $f_\infty$ of $f_n$ (recall \eqref{psi-n-def} and \eqref{psi-inf-def})
\begin{equation}
f_\infty(x_0)\,=\,\psi_\infty (x_0)+ \frac{\log b}{p-1}\, .
\end{equation}
 Note that, while from the construction { of the lattice 
$b=2,3 , \ldots$  is  an integer}, the iteration { (\ref{HL})}  makes sense for every $b>1$. 
{ In what follows   we will consider this generalization} (see Remark~\ref{rem:diamond-GW} for { another probabilistic interpretation  of
} $Z_n$ in this generalized context).
}
Finding the critical behavior of the free energy when the temperature is close to the wetting transition temperature, i.e. 
when $x_0$ is close to the fixed point $x_*$ of the map $F$,  is a problem of the same nature as the one posed in the introduction and
 the question we address here is to try to  determine the periodic function $B(\cdot)$  in front of the power law singularity
\begin{equation}
\label{eq:B}
f_\infty(x)-{\log b \over p-1} \stackrel{x \searrow x_*}\sim 
  (x-x_*)^\alpha \  B\left(-{\log(x-x_*) \over \log F'(x_*) }\right)\, . 
\end{equation}

The map (\ref{HL}){ can be reduced to the form   (\ref{F-def}) or \eqref{large-p2} 
in the two following cases}
\begin{itemize}
\item  when $b$ is close to $1$ 
or  when $b$ is  large,  by making the change of variable $x = b^{1 \over p-1} X$,  the map (\ref{HL})  becomes
\begin{equation}X \to X^p +{ b^{-{1 \over p-1}}  -  b^{-{p \over p-1}} }\, ,  \end{equation}
which is  of the form (\ref{F-def}). 
\item when $b$ is arbitrary and $p$ is large, because the application { (\ref{HL})} coincides with  \eqref{large-p2},
with the choice $q=1/b$ and $P(\cdot)\equiv 1$. 
\end{itemize}     


\begin{rem}
\label{rem:nb}
\rm
It is worth pointing out that the partition function 
of the wetting model on more general  hierarchical lattices  is obtained by iterating a polynomial  of 
the form $F(x)=q_0+q_1x+\dots +q_p x^p$, with the $q_j$'s probability weights. Of course $F(1)=1$ and it is easy to see
that there is at most another positive root: recall that the unstable fixed point $x_*$ on which we focus is the largest positive root. 
We record here the well known formula (e.g. \cite{cf:Harris,wet1,cf:BB}) that actually holds for every $x \in (0, \infty)$
\begin{equation}
\label{eq:F.1}
\psi_\infty(x)\, =\, \frac{\log  q_p}{p-1}+ \log x+ 
\sum_{i=0 }^\infty  p^{-(i+1)}{Q\left(F^{(i)}(x)\right)}\,, 
\end{equation}
with $Q(x):=\log( F(x)x^{-p}/q_p)$. It is rather straightforward to extract from this formula that
$\psi_\infty(\cdot)$ is (real) analytic on $(x_*, \infty)$ (e.g. \cite{cf:CG}): in statistical mechanics terms this says that
$x_*$ is the only critical point (note that { with the definition (\ref{psi-n-def}, \ref{psi-inf-def})} one has $\psi_\infty(x)=0$ for $x \in (0,x_*]$).
\end{rem}

\subsection{The Galton Watson process}
The Galton process is a simple model for the  size of  the lineage of an individual \cite{cf:HarrisBook}.
The generations do not overlap and all the population is replaced at every generation.

One starts with a single  individual at generation $g=0$.  Then    each individual living at generation $g$  has a probability $q_k$ of having
$k$ offspring at the next generation $g+1$. Therefore if $N_g$ is the number of  individuals at generation $g$ (so $N_0=1$), one has
\begin{equation}
N_{g+1} \, =\,  \sum_{i=1}^{N_{g}}  k_i\, ,
\label{recurs}
\end{equation}
where the $k_i$'s are $N_{g}$ independent random variables which take the value  $k$ with a probability  $q_k$.
The size  $N_g$ of the population at generation $g$ is therefore a random variable. 
The average number of offsprings at generation $g$ is  (see (\ref{recurs})) given by
\begin{equation}
\langle 
N_g \rangle \, =\,   \left( \sum_{k} k \, q_k \right)^g \, =\, \langle k \rangle^g\, .
\end{equation}
So the expected size of the population grows exponentially for $\langle k\rangle >1$. It is well known, see e.g.
\cite{cf:HarrisBook}, that if $\langle k\rangle \le 1$ then after sufficiently many generations the population goes extinct.
We will actually restrict to the {\sl supercritical} case $\langle k\rangle >1$ and in this context it  is natural to define 
\cite{cf:Harris,cf:HarrisBook}
 the  generating function  of the ratio  $N_g/ \langle N_g \rangle$ 
\begin{equation} 
H_g(\lambda) \, :=\,   \left\langle \exp\left(\lambda {N_g \over \langle N_g \rangle} \right)  \right\rangle\, ,
\end{equation}
One can see  from (\ref{recurs}) that
\begin{equation}
H_{g}(\lambda) \,  =\,   \sum_k  \left(H_{g-1}\left({\lambda \over \langle k \rangle}\right)\right)^k \,  q_k\,=\,
 F\left(H_{g-1}\left({\lambda \over \langle k \rangle}  \right) \right)
\, =\, F^{(g)} \left(H_0\left({\lambda \over \langle k \rangle^g}  \right) \right)\, ,
\end{equation}
where $F(x)= \sum_k q_k  \ x^k$.
Then  $H_{\infty}(\lambda): =\lim_{g \to \infty}H_{g}(\lambda)$   satisfies
\begin{equation}
\label{eq:Poincare}
H_{\infty}(\lambda)\, =\, F\left(H_{\infty}\left({\lambda \over \langle k \rangle}  \right) \right)\, ,
 \end{equation}
 This functional equation is known under the name of Poincar\'e equation  \cite{cf:Milnor}
 and $H_\infty(\cdot)$ is the unique solution
 of  \eqref{eq:Poincare} which is analytic in a neighborhood of the origin and such that
 $H_\infty(0)=1$ and $H'_\infty (0)=1$
 \cite{cf:Milnor}. Note that that, {given $H_\infty(0)$ and $H'_\infty (0)$, 
 one can extract recursively
from (\ref{eq:Poincare}) }
 all the coefficients of the Taylor expansion at $\lambda=0$.

\medskip

\begin{rem}
\label{rem:W}
\rm 
To be precise, the limit exists in great generality and it is (obviously) finite if $\lambda \le 0$
\cite[Ch. I]{cf:HarrisBook}. In fact the random variable $W_g:= N_g/\langle N_g \rangle$
converges almost surely to a limit random variable $W_\infty$ as soon as $\langle  k \rangle <\infty$:
this is particularly elementary if $\langle  k^2 \rangle <\infty$, because in this case one
can explicitly compute $\langle (W_{g+n}-W_{g})^2\rangle$ for $n\in \mathbb{N}$
\cite[p.~13]{cf:HarrisBook} and establish the convergence (in probability) of $W_g$ to $W_\infty$ and
the fact that $W_\infty$ is not identically zero.
Of course in this generality $H_{\infty}(\lambda)$ can be $\infty$ for $\lambda>0$, but we will focus on
 the case
 in which  only a finite number of $q_k$'s are non zero and 
  it is therefore  rather straightforward to see 
that for every $\lambda>0$ there exists $C_\lambda$ such that
\begin{equation}
\sup_g \left \langle \exp( \lambda W_g ) \right\rangle \, \le \, C_\lambda\, .
\end{equation}
This not only implies that $H_\infty(\lambda)$   is finite for every $\lambda\in \mathbb{R}$
and that it  is equal to $\langle \exp( \lambda W_\infty ) \rangle$, but also that the same is true for any $\lambda$ complex: as  a matter of fact
 $H_\infty(\cdot)$ is an entire function, that is analytic on the whole of
$\mathbb{C}$. 
\end{rem}

\medskip

Let us therefore assume that only a finite number of $q_k$'s are non zero, so that  the map $F(x)$ is polynomial  of degree $p=\max\{k:\, q_k \neq 0\}$.
{T. E. Harris} has proven \cite{cf:Harris} that for  large $\lambda$ 
\begin{equation}
\label{Harris}
\log H_\infty(\lambda) \sim \lambda^\alpha L\left( {\log \lambda \over \log \langle k \rangle} \right) \, ,
\end{equation}
with $\alpha={\log p/ \log \langle k \rangle} $ { as in (\ref{alpha})} 
and  $L(\cdot)$  a continuous positive  periodic function of period $1$. Harris was unable to establish that
$L(\cdot)$ is not a constant as soon as $q_p<1$ (for $q_p=1$, that is for $F(x)=x^p$, it is straightforward to see that $L(\cdot)$ is constant) 
 and this open issue has drawn the attention in the mathematical community (see for example \cite{cf:BB,cf:BN}, dealing precisely 
with the problem left open by Harris) and similar -- sometimes strictly related -- oscillatory behaviors have been pointed out repeatedly (e.g. \cite{cf:KMcG,cf:dubuc,cf:Odl,cf:GW,cf:teufl} and \cite{cf:CG} for further references: the stress is often on the {\sl nearly constant} and {\sl nearly sinusoidal} character of these oscillations). 
To our knowledge, establishing in general that $L(\cdot)$ is non constant is  still an open problem for $p>2$
(for $p=2$ the non triviality of $L(\cdot)$ is established in \cite{cf:CG} by exploiting results in \cite{cf:CH2}).

\medskip

Here again one can try to perturb around a situation  where the map becomes simple. However there is the following important
preliminary issue: what is the relation, if any, between the periodic functions $B(\cdot)$ and 
$L(\cdot)$, both of period one, appearing in \eqref{eq:B} and  
in \eqref{Harris}. A priori, these two periodic function come out of different questions that have in common only
the underlying polynomial $F$. But in fact if the polynomial is the same in the two questions, then the periodic functions
coincide up to the change of variable {$ B(y)=L(-y)$}   (this has been proven in \cite{cf:CG} and the need for switching the sign
is due to the way we have defined $B(\cdot)$ and 
$L(\cdot)$: the argument of proof is reproduced below). Given this observation,
the analysis of $B(\cdot)$ and 
$L(\cdot)$ in the two perturbative limits we consider becomes just one problem.  

\subsection{Oscillations for hierarchical wetting and Galton-Watson models} 
We explain now how,
by exploiting the two functional equations \eqref{eq-fonct} and 
\eqref{eq:Poincare},
one can see that the periodic function $B(\cdot)$ in \eqref{eq:B} and $L(\cdot)$
in \eqref{Harris}  coincide up to a sign change in the argument. For this we fix a polynomial $F(x)=\sum_{i=0}^p q_i x^i$,
like in Remark~\ref{rem:nb} for the wetting model  and in Remark~\ref{rem:W} for the Galton-Watson process,  and we
{ define}  for every $\lambda>0$
\begin{equation}
\label{eq:key}
M(\lambda)\, =\, \lambda^{-\alpha} \psi_\infty \left( H _\infty ( \lambda ) \right)\, ,
\end{equation}
and by applying first 
\eqref{eq:Poincare} together with $\alpha=\log p/ \log \langle k \rangle$, and then \eqref{eq-fonct} 
we have
\begin{equation}
M\left( \lambda \langle k \rangle\right)\,=\, \frac{\lambda^{-\alpha}}{p} 
\psi_\infty \left( F\left(H _\infty ( \lambda )\right) \right)\, =\, 
\lambda^{-\alpha}
\psi_\infty \left( H _\infty ( \lambda ) \right)\, =\, M(\lambda)\,,
\end{equation}
that is, $M(\cdot)$ is multiplicatively periodic of period $\langle k \rangle$.
Moreover $M(\cdot)$ is analytic on $(0, \infty)$  because
$H_\infty(\cdot)$ is entire (cf. Remark~\ref{rem:W}) and $\psi_\infty(\cdot)$ is analytic on $(1, \infty)$ (cf. Remark~\ref{rem:nb}).  
The fact that 
$M(\cdot)>0$ on $(0, \infty)$ is just 
the strict positivity of $\psi_\infty(x)$ for $x>1$ and the fact that $H_\infty(\lambda)>1$ for $\lambda>0$.

{ Now  from the definitions it is} immediate to see that $\psi_\infty(x) \sim \log x$ for $x \to \infty$
and that $\lim_{\lambda \to \infty} H_\infty (\lambda)=\infty$,
so  \eqref{eq:key} 
\begin{equation}
\label{eq:pHarris}
\log \left( H_\infty (\lambda) \right) \stackrel{\lambda \to \infty}{\sim} \lambda^\alpha M(\lambda)\, ,
\end{equation}
which provides
a proof of  \eqref{Harris} and relates explicitly $M(\cdot)$ and $L(\cdot)$.

On the other hand \eqref{eq:key} can be rewritten as
\begin{equation}
\label{eq:fkey}
\psi_\infty (x) \, =\, \left(H_\infty^{-1}(x) \right)^\alpha
M\left(H_\infty^{-1}(x)
\right)\, .
\end{equation} 
But since
$H_\infty(0)=1$ and $H'_\infty(0)=1$, we see that
$H_\infty^{-1}(1+y) \sim  y$ for $y\searrow 0$, so that
from \eqref{eq:fkey} one directly obtains for $x\searrow 1$
  \begin{equation}
\psi_\infty (x) \, \sim \, \left(x-1 \right)^\alpha
M\left(x-1
\right)\, ,
\end{equation} 
which proves \eqref{eq:B} and explicitly relates $B(\cdot)$ to $M(\cdot)$, and we see that 
{ $B(y)=L(-y)$ for every $y$.}
This completes the proof of the equivalence of the two example models we have proposed, as far as
oscillations are concerned.
}

\medskip
\begin{rem}
\label{rem:diamond-GW}
\rm
As it was pointed out in \cite{wet3}, 
the partition function of the hierarchical wetting model has a Galton-Watson representation:
\begin{equation}
\label{eq:hGW}
\frac{Z_n(T)}{Z_n(\infty)}
 \, =\, \left \langle \exp \left(\frac a T  N_n \right) \right\rangle\, .
\end{equation}
 Formula \eqref{eq:hGW} shows that { the partition function of } the hierarchical wetting model can be interpreted  { as the generating function of the size $N_n$ of the population at the $n$-th generation of
} a  Galton-Watson processes.
It also provides an expression for the partition function of the hierarchical model in terms of sum of Boltzmann weights  when $b$
is not integer. 
\end{rem}


\section{Main results}
\label{sec:main}
\subsection{ Oscillations for the  map  (\ref{F-def}-\ref{G1}) for   $\epsilon$ small}

Here instead of writing the results in terms of the difference $x-x_*$, we found slightly  more convenient to use $\log(x/x_*)$.  One can of course  make a simple change of variables to rewrite  the final result in terms of the difference  $x-x_*$.
Then (\ref{crit-1}) becomes for $x$ close to $x_*$
\begin{equation}
\psi_\infty(x) \   \simeq  \   
\left(\log[x / x_*] \right)^\alpha B\left( -{\log  \left(\log[x / x_*] \right) \over \log F'(x_*)} \right)  \, ,
\end{equation}
 where $B(\cdot)$ is a periodic function.
By replacing $\psi_\infty(x)$ by this leading behavior in (\ref{eq-fonct}) one can calculate subdominant contributions when $x \to x_*$ and one gets at next order
\begin{eqnarray}
\psi_\infty(x) &\simeq & 
\left(\log[x / x_*] \right)^\alpha B\left( -{\log  \left(\log[x / x_*] \right) \over \log F'(x_*)} \right)  
\nonumber
 \\ &&  - \left(\log[x / x_*] \right)^{\alpha +1}{ \alpha  [ x_*    F''(x_*)  + F'(x_*) - F'(x_*)^2]\over 2 F'(x_*)(F'(x_*) -1) }
\label{crit-3}
\\ && \ \ \ \ \ \ \ \ \ \ \ \ \  \times  \left[
B\left(- {\log  \left(\log[x / x_*] \right) \over \log F'(x_*)} \right) - {1 \over \log p} B'\left( -{\log  \left(\log[x / x_* ]\right) \over \log F'(x_*)} \right)  \right]  + ... 
\nonumber
\end{eqnarray}
In fact all the subdominant contributions can be expressed in terms of derivatives of the map $F(x)$  evaluated at  $x_*$ and of derivatives of the periodic function $B(y)$. So the problem of calculating the amplitudes of the dominant contribution and of all subdominant contributions reduces to the calculation of the periodic function $B(y)$.

\medskip 

One of our main results, derived below under the condition \eqref{G1}  is that
\begin{equation}
B(y)\,=\, 1 + \epsilon B_1(y)+ \epsilon^2 B_2(y) + \ldots
\end{equation}
 with
\begin{eqnarray}
B_1(y) \, =\,  {1 \over p}  \left[-  y \, G'(1)   +  \sum_{k=0}^\infty p^{y-k} G \left( e^{p^{k-y}} \right)  + \sum_{k=-\infty }^{-1}\left[ p^{y-k}
G \left( e^{p^{k-y}} \right)  -G'(1) \right] \right]\, , 
\label{B1-def}
\end{eqnarray}
and 
\begin{multline}
B_2(y) 
\, =\,     {y(y+1) \over 2 p^2}\, G'(1)^2  
\\
 +{ 1  \over p}
 \Bigg[ 
 \sum_{k \ge 0} 
 \left(
-{ p^{y-k} \over 2}  G^2\left( e^{p^{k-y}} \right)  + e^{p^{k-y}}  G' \left( e^{p^{k-y}} \right) \left[ B_1(y) -  p^{y-k}  U  \left( e^{p^{k-y}} \right) \right] 
\right)
  \\
+ \sum_{k =-\infty}^{-1} 
\bigg(  -{G'(1)^2 \over p} (k-y) 
- {p^{y-k} \over 2}  G^2\left( e^{p^{k-y}} \right)  
\\
+  e^{p^{k-y}}  G' \left( e^{p^{k-y}} \right) \left[ B_1(y) -  p^{y-k}  U  \left( e^{p^{k-y}} \right) \right] \bigg)
 \Bigg]\, ,
\label{B2-def}
\end{multline}
where $U(x)$ is defined by 
\begin{equation}
\label{U-def} 
U(x)= {1 \over p}\sum_{k=0}^{\infty} p^{-k} G\left(x^{p^k}\right) \, .
\end{equation}
\medskip

\begin{rem}
\rm 
\label{rem:pc}
In the particular case
\begin{equation}
F(x)\,=\, (1-\epsilon) \, x^p + \epsilon\, 
\end{equation}
one has $G(x)=x^{-p}-1$
and therefore 
{
\begin{equation}
\label{eq:B1part}
B_1(y)\, =\, y + \sum_{k=1}^\infty g(y-k)+
\sum_{k=-\infty}^0 \left(g(y-k)+1 \right)\, ,
\end{equation}
with
\begin{equation}
g(y)\, =\, p^y \left( \exp(-p^{-y})-1 \right)\, .
\end{equation}
From this expression one can obtain   the Fourier coefficients $a_n = \int_0^1 \exp(-2\pi n i y) B_1(y) \mathrm{d} y$
of the periodic function $B_1(\cdot)$}
and one finds
\begin{equation}
\label{eq:fB1part-0}
a_0\,=\,   {1 \over 2} -
 {1+ \Gamma'(1) \over \log p}\, =\, {1 \over 2} -
 {1-\gamma \over \log p}\, ,
\end{equation}
($\gamma=0.5772\ldots$ is the Euler-Mascheroni constant)
 and  for $n \neq 0$ 
\begin{equation}
\label{eq:fB1part-n}
a_n\, =\,  {1 \over \log p} \, \Gamma \left({2 i \pi n  \over \log p} -1\right) \,  .
\end{equation}
{ These espressions are compared in Table 1 with 
the Fourier coefficients  calculated directly for a finite value of $\epsilon$.}
\end{rem}

\medskip
\begin{center}
\begin{table}[h]
\small
   \begin{tabular}{| l |  | l |  l  | l | l  | l | }
     \hline
     &  &  & & &     \ \\[-11pt]
     $1/\epsilon$ & $(a_0-1)/\epsilon$  & $\Re(a_1/\epsilon){ \times} 10^8$ &  $\Im (a_1/\epsilon){\times} 10^{8}$ & $\Re(a_2/\epsilon){\times  }10^{14}$ &  $\Im (a_2/\epsilon){ \times }10^{14}$\\ \hline \hline
     $10$ & \ 0.129 & $-3.65$ & $-7.08$&$1.47 $& $-0.80 $\\    
     $10^2$ &  -0.093 & $-4.80 $& $-7.07$ & $1.62$ & $-1.11$ \\ 
     $10^3$ & -0.108& $-4.788$&$-7.164$ & $1.682$ & $-1.085$\\ 
     $10^4$ & -0.0978 & $-4.786$& $-7.174$ & $1.6887$ &$-1.0819$\\ 
     $10^5$ & -0.10993 & $-4.7858$&$-7.1753$ & $1.68938$ &$-1.08166$ \\  \hline
     $\infty$ & -0.10995  & $-4.7858$& $-7.1757$ & $1.689459$ & $-1.081618$\\ \hline
   \end{tabular}
   \medskip
\caption{\footnotesize  For $F(x)= (1-\epsilon) x^2+ \epsilon$ the limit $\epsilon \searrow 0$ expressions  in \eqref{eq:fB1part-0} and in \eqref{eq:fB1part-n} are compared { with 
 values for $\epsilon>0$  obtained by using formula \eqref{eq:key}:} an accurate expression for $H_\infty(\cdot)$ is obtained by writing enough terms in the Taylor expansion at zero (these terms are easily
obtained recursively by using $H_\infty(0)=1$, $H'_\infty(0)=1$ and by taking derivatives of the  Poincar\'e equation \eqref{eq:Poincare}) and by exploiting the series in \eqref{eq:F.1} to get a good approximation of $\psi_\infty(\cdot)$.
}
   \end{table}
 \end{center}

\medskip

\begin{rem} \rm
\label{rem:Poisson} 
\eqref{eq:fB1part-0} and \eqref{eq:fB1part-n} follow from 
\eqref{eq:B1part} by direct computation. It is however worthwhile to point out that there is a direct  link with a Poisson summation formula (and, for example, with a  similar computation in 
a different context \cite[\S~3.3]{cf:Sornette}). In fact, we have
\begin{equation}
\label{eq:B1part2}
B'_1(y)\, =\, 1 + \sum_{k=-\infty}^\infty g'(y+k)\, ,
\end{equation}
so the $n$-th Fourier coefficient of $B'_1(\cdot)$ is equal to
 $\int_{-\infty}^\infty g' (y) \exp(-2\pi i n y ) \text{d} y=:
 \widehat{g'}(2\pi n)$ for $ n\neq 0$ ({\sl Poisson summation formula}, see e.g. \cite[Ch.~4]{cf:Pinsky}), while the $0$-th Fourier coefficient is of course $0$.
 It is then straightforward to see that $a_n =  i \widehat{g'}(2\pi n) / {2\pi n}$ and 
 \eqref{eq:fB1part-n} can be recovered this way too.
 \end{rem}

\subsection{Large $p$ limit of the map (\ref{large-p2})}
 For the map \eqref{large-p2}, for $q$ and $P(\cdot)$ fixed and 
$p$   large, so that in particular $F'(1)= qp +(1-q) P'(1)\sim qp$, 
 our  result is that  
if \begin{equation} - {\log( x-1) \over \log(F'(1)) } = n(x) + y \ \  \ \ \ {\rm with} \ \ \ \  \eta < y < \eta+1
\, ,
\end{equation}
where $n(x)$ is an integer, 
then for $y\neq 1$  as $p \to \infty$
\begin{equation}
\label{lp}
\psi_\infty(x) \simeq \begin{cases}
  (x-1)^\alpha   \,    q^{-y}  &  \ \ \ {\rm  for}   \ \  \eta< y < 1 \, ,
\\  
 (x-1)^\alpha\,  q^{1-y} &   \ \ \ {\rm  for}  \ \  1<y < 1+\eta \, ,
 \end{cases}
\end{equation}
This gives a discontinuous amplitude at $y=1$. Recalling \eqref{crit-1},
 an equivalent way  to state \eqref{lp}
is to say that for every   $y \not\in \mathbb{Z}$
\begin{equation}
\lim_{p \to \infty} B(y) \, =\, q ^{-\{ y\}}\, ,
\label{B-p-inf}
\end{equation}
where $\{ x\}$ is the fractionary part of $x$, that is $ \{ x\}= x-\max\{n\in \mathbb{Z}:\, n\le x\}$.  
\medskip

\begin{figure}[hlt]
\begin{center}
\leavevmode
\epsfxsize =14 cm
\psfragscanon
\psfrag{A}{\large $y$}
\psfrag{B}{\large $B(y)$}
\epsfbox{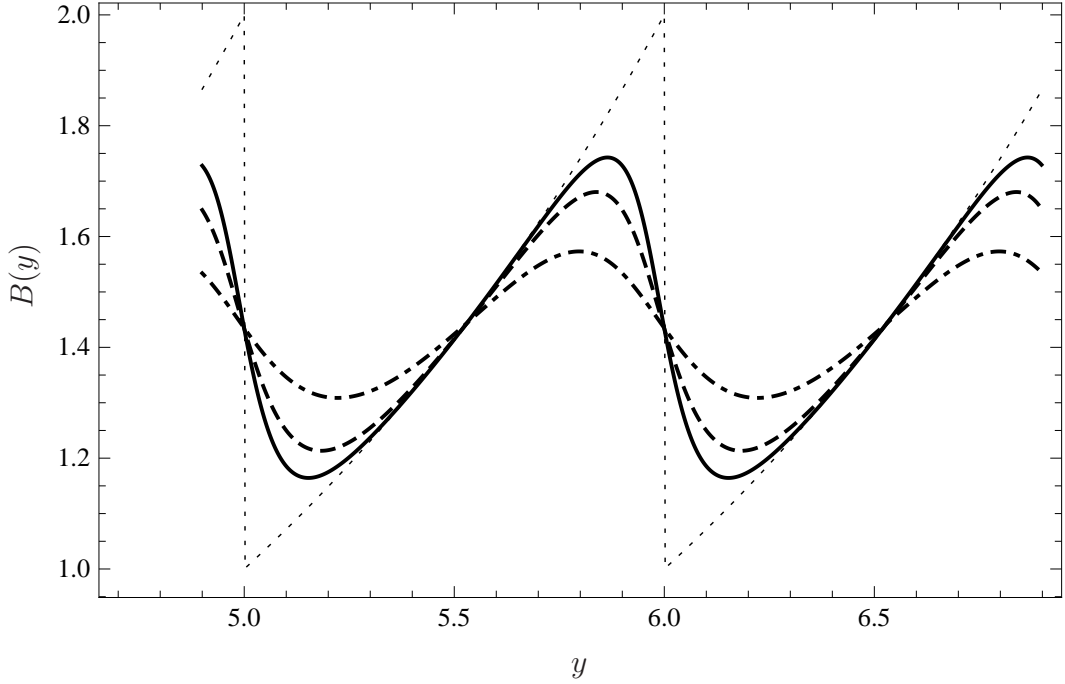}
\end{center}
\caption{ With the choice $F(x)=(x^p+1)/2$ we have plotted $B(\cdot)$ for
$p=10^3$ (dot-dashed line), $10^5$ (dashed line) and $10^7$ (solid line),   and the this dotted line for  $p=\infty$. For the $p=\infty$ case we have exploited the result (\ref{B-p-inf}),   that is we have plotted $y\mapsto 2^{\{y\}}$.
The numerical data for $p<\infty$ have been obtained  exploiting \eqref{eq:F.1}. The convergence
is rather slow simply because the effective large parameter is $\log p$. Notice that already at $p=1000$
$B(\cdot)$ is far from being {\sl nearly constant} and starting from $p=10^5$ it is also evident that it is not nearly sinusoidal.
These large amplitudes contrast with  the small amplitudes one
 usually observes for small $p$ \cite{cf:BB,cf:BN,cal0,cf:CG}.}
\label{fig:1}
\end{figure}

One can resolve the discontinuity at $y=0$ by analyzing, in the large $p$ limit, the range
$y-1 \sim (\log p)^{-1}$
and one gets
\begin{equation}
 \psi_\infty(x) \simeq  (x-1)^\alpha  \   e^{-(1-y)\log p} \  \log
\left[q  \exp\left( {e^{(1-y) \log p } \over q} \right)  + 1-q \right]    
\label{resolve}
\end{equation}
Note also that in the case $q=1-\epsilon$ and $P(\cdot)\equiv 1$
 for large $p$ one gets  from \eqref{B1-def}-\eqref{B2-def}   
  in the range  $0< y < 1$
\begin{equation}
B_1(y)\, =\, y \ \ \ \text{ and } 
 \ \ \ B_2(y)\, =\,  {y (y+1) \over 2}\, ,
  \end{equation}
which agrees with \eqref{lp} when one takes the first two terms {in the expansion in powers  of} $\epsilon$.

\section{  Perturbation in powers of $\epsilon$ far from the critical point}
\label{sec:perturb}
Our approach consists  first  in calculating   in this section $\psi_\infty(x)$ in powers of $\epsilon$,    for $x \neq x_*$.
Then by matching  (in Section \ref{sec:expect}) the singularities (obtained in Section \ref{sec:sing}) of the successive terms of the expansion with the expected expansion (see Section  \ref{sec:expect}) near the critical point,  one gets the periodic functions $B_1$ and $B_2$.
\medskip

Starting from the map (\ref{F-def}) 
and defining the successive terms of the expansion of  $F^{(n)}(x)$  in powers of $\epsilon$ by 
\begin{equation}
F^{(n)}(x)\, =\,  x^{p^n} \exp\left(\epsilon  p^n Y_n(x) + \epsilon^2 p^n  Z_n(x) + O(\epsilon^3) \right)\, .
 \end{equation}
 Then by replacing in the expression (\ref{F-def}) one gets
 \begin{equation}
 \begin{split}
 Y_n(x)&= Y_{n-1}(x) + p^{-n} G\left(x^{p^{n-1}}\right)\, ,
  \\
 Z_n(x)&= Z_{n-1}(x) -{1 \over 2\, p^{n}} G^2\left(x^{p^{n-1}}\right) 
+{1 \over p} \, G'\left(x^{p^{n-1}}\right)  \,  x^{p^{n-1}}  \, Y_{n-1}(x)\ . 
\end{split}
\end{equation}
One can show by recursion that
\begin{equation}Y_n(x) ={1 \over p} \sum_{k=0}^{n-1} p^{-k} \,  G \left( x^{p^k} \right) = U(x) - {U \left( x^{p^n} \right)  \over p^{n} }\, ,\end{equation}
where  we recall $U(x)$ from (\ref{U-def})
\begin{equation}
U(x)={1 \over p}\sum_{k=0}^{\infty} p^{-k} G\left(x^{p^k}\right) \, ,
\end{equation}
and this  leads to
\begin{equation}
\begin{split}
\psi_n(x)= &  \,  \log x + \epsilon  \,   Y_n(x) + \epsilon^2  \,   Z_n(x) + O(\epsilon^3)
\\
= &   \,  \log x + \epsilon  \left(   U(x) - {U \left( x^{p^n} \right)  \over p^{n} } \right) 
   \ - \ {\epsilon^2 \over 2  \, p } \, \sum_{k=0}^{n-1} p^{-k} \, G\left(x^{p^k}\right)^2
 \\  & \, \,
+{\epsilon^2  \over p} \, \sum_{k=0}^{n-1} x^{p^k} \, G'\left(x^{p^k}\right)  \left(   U(x) - {U \left( x^{p^k} \right)  \over p^{k} } \right) 
  \  + \   O (\epsilon^3)\, .
  \end{split}
\end{equation}
For large $n$, assuming $G(\cdot)$ is such that the sums converge, one gets
\begin{equation}
\psi_\infty(x) = \log x + \epsilon   \ U(x) +  \epsilon^2 \left(-  {V_1(x)  \over 2} +   U(x) \,  V_2(x) -  V_3(x)\right) + ... 
\label{psi-inf}
\end{equation}
 where
\begin{equation}
\label{V-def} 
\begin{split}
V_1(x)  &\, =\, {1 \over p}  \sum_{k=0}^{\infty} p^{-k} G^2\left(x^{p^k}\right)\, ,
\\
V_2(x)&\, =\, {1 \over p} \sum_{k=0}^{\infty}x^{p^k}   \  G'\left(x^{p^k}\right) \, ,
\\
V_3(x)&\, =\,  {1 \over p}\sum_{k=0}^{\infty} p^{-k} x^{p^k}   \  G'\left(x^{p^k}\right) \ U\left(x^{p^k}\right)\, . 
\end{split}
\end{equation}
Since $G(\cdot)$ is bounded, the first series converges. Moreover if the series for $V_2(\cdot)$
converges, so does the series defining $V_3(\cdot)$. Therefore we require 
the converges of the series defining $V_2(\cdot)$ and this amounts to an additional assumption 
on the large $x$ behavior of $G'(x)$: for example, it suffices that $\vert G'(x)\vert \le 1/x(\log x)^a)$ for some $a>1$
and $x$ large.

\section{Singular behavior of the  sums \eqref{U-def}-\eqref{V-def}}
\label{sec:sing}
In this section we try to extract the singular behavior, as $x \searrow 1$,
 of the sums $U(x), V_1(x), V_2(x)$ and $V_3(x)$ defined in \eqref{U-def}-\eqref{V-def}.
\medskip 

For $x$ sufficiently close to $1$ one can write  
\begin{equation}
\log(\log x)\,  =\,  -(n_0 + y) \log p\, ,
\label{n0-def}
\end{equation}
$n_0$ is an integer
where  $0<y<1$. Then as $x\searrow 1$, the integer $ 
  n_0 \to \infty.$

We will also assume that as in (\ref{G1}), one has $G(1)=0$,
and since $G(\cdot)$ is smooth 
\begin{equation}
G(x) = G'(1) (x-1) + G''(1) {(x-1)^2 \over 2} + ...
\label{G1-exp}
\end{equation}

\subsection{Analysis of $U(x)$ as $x \searrow 1$}  
One can easily see, { using $G(1)=0$ and the definition (\ref{n0-def}) of $n_0$}, that $U(x)$ defined in (\ref{U-def}) can be written as 
\begin{multline}
U(x)  \, 
= \, {\log x  \over p} \left( \sum_{k=0}^\infty p^{y-k} G \left( e^{p^{k-y}} \right)  + \sum_{k=-n_0}^{-1} \left[ p^{y-k} G \left( e^{p^{k-y}} \right)  -G'(1)\right]   \right)  \\ + {\log x  \over p} \left( - {\log(\log x) \over \log p} - y \right) G'(1) \, ,
\end{multline}
{Then, by using (\ref{G1-exp}), we see that
 $n_0 \to \infty$ as $x \searrow 1$ and}
\begin{equation}U(x) - U^*(x) = O \left(\log^2 x \right)\, ,
 \end{equation}
where  $U^*(x)$ is defined as
\begin{equation}
 U^*(x)=  - {G'(1) \over p \log p} (\log x ) (\log(\log x)) + B_1(y) \log x \, ,
\label{Ustar}
\end{equation}
 where $B_1(\cdot)$ is  the periodic function ($B_1(y)=B_1(y+1)$)  defined in  (\ref{B1-def}).
 

\medskip
A few remarks:
\begin{enumerate}
\item 
 One can show that
\begin{equation}U(x)= {1 \over p} U(x^p) + {1 \over p} G(x)\, ,
\end{equation}
\begin{equation}U^*(x)= {1 \over p} U^*(x^p)   + {1 \over p} G'(1) \log x\, .\end{equation}
\item 
One can also evaluate the correction $U(x)-U^*(x)$ 
 and  get for $x \searrow 1$
\begin{equation}U(x)= U^*(x) - {G''(1)+G'(1) \over 2 (p-1)p} \log^2(x) + O\left(\log^3 (x)\right)\, .
 \end{equation}
\item
One can check directly from the full expression (\ref{B1-def}) of $B_1(y)$ that
\begin{equation}B_1(y+1)\, =\, B_1(y)\, .\end{equation}
\end{enumerate}
\subsection{                Analysis of $V_1(x)$ as $x \searrow 1$} 
Using the definition (\ref{n0-def}) of $n_0$   and the behavior (\ref{G1-exp}) of $G(x)$ when $x \searrow 1$ one can rewrite the sum in the first line of \eqref{V-def} as
\begin{eqnarray}
V_1(x) (x)  
= {\log x  \over p}  \sum_{k=-n_0}^\infty p^{y-k} G^2 \left( e^{p^{k-y}} \right)\ ,
\end{eqnarray}
and as $n_0 \to \infty$  in the limit $x \searrow 1$ one gets
\begin{equation}
\label{V1-result}
 V_1(x)   \simeq  {C_1}(y) \log x\ ,
\end{equation}
where $C_1(\cdot)$ is the periodic function ($C_1(y+1)=C_1(y)$)  defined by
\begin{eqnarray}
{C_1}(y) = {1 \over p}  \sum_{k=-\infty }^\infty p^{y-k} G^2 \left( e^{p^{k-y}} \right) \, .
\label{C1-result}
\end{eqnarray}
\medskip

\subsection{Analysis of $V_2(x)$ as $x \searrow 1$} 
Using again the relation (\ref{n0-def})  between $x$ and $n_0$ one can write the sum in the second expression in \eqref{V-def} 
\begin{multline}
 V_2(x)
= {1 \over p} \left[ \sum_{k=0}^\infty e^{p^{k-y}}  G' \left( e^{p^{k-y}} \right)  + \sum_{k=-n_0}^{-1} e^{p^{k-y}}  G' \left( e^{p^{k-y}}\right) - G'(1) \right]  + { G'(1)   \over p} n_0  \, \simeq \\
{1 \over p}\left[ \sum_{k=0}^\infty e^{p^{k-y}}  G' \left( e^{p^{k-y}} \right)  + \sum_{k=-\infty}^{-1} \left[ e^{p^{k-y}}  G' \left( e^{p^{k-y}}\right) - G'(1) \right]  \right] 
   - {\log(\log x)   \over p \log p } G'(1)  - {G'(1)  \over p}  y\, .
\end{multline}
One then finds
\begin{equation} 
V_2(x)= -G'(1) {\log(\log x) \over p \log p} + C_2(y)  + O(\log x) \, ,
\label{V2-result}
\end{equation}
with with the periodic function $C_2(\cdot)$ given by
\begin{equation} 
C_2(y) =  {1 \over p} \left[- y  \, G'(1)  +\sum_{k=0}^\infty e^{p^{k-y}} G' \left( e^{p^{k-y}} \right)  + \sum_{k=-\infty }^{-1} \left[ e^{p^{k-y}} G' \left( e^{p^{k-y}} \right) - G'(1) \right] \right] 
\label{C2-result}\,.
\end{equation}

\medskip

\begin{rem}\rm
 One can easily see   from the definitions \eqref{U-def} and \eqref{V-def} of $U(x)$ and $V_2(x)$ that
\begin{equation}V_2(x) \, =\,  x  \, U'(x)\, .\end{equation}
This implies that
\begin{equation} 
             C_2 (y)= B_1(y) - {B_1'(y) \over \log p} - {G'(1) \over  p  \ \log p} \, ,
\label{C2-result-bis}
\end{equation}
which can be also checked directly.
\end{rem}
\medskip

\subsection{ Analysis of $V_3(x)$ as $x \searrow 1$}  
From the definitions  in the third expression in \eqref{V-def} and \eqref{n0-def}  of the sum $V_3$  and of $n_0$ one can write
\begin{eqnarray}
V_3(x)
=   & & {\log x  \over p}   
\sum_{k \ge 0} p^{-k + n_0 + y} 
 e^{p^{k-n_0-y}}  G' \left( e^{p^{k-n_0-y}} \right)  U  \left( e^{p^{k-n_0-y}} \right)  \, ,
\end{eqnarray}
which can be rewritten  using again (\ref{n0-def}) 
\begin{equation}
\begin{split}
V_3(x)
=\,   & { \log x  \over p}
\left[ \sum_{k \ge 0} p^{-k  + y}
 e^{p^{k-y}}  G' \left( e^{p^{k-y}} \right)  U  \left( e^{p^{k-y}} \right) \right.  \\
  &  \left.
+ \sum_{k \ge -n_0}^{-1} \left\{ p^{-k  + y}
  e^{p^{k-y}}  G' \left( e^{p^{k-y}} \right)  U  \left( e^{p^{k-y}} \right)    
   +{G'(1)^2 \over p } (k-y)   - B_1(y) G'(1)  \right\}  \right.
\\
 & \left.  +{G'(1)^2 \over 2  p }   \left( {-\log( \log x)  \over \log p} - y \right)  \left( {-\log( \log x)  \over \log p} - y +1\right)
\right.  \\ 
  &  \left.+  \left(  {y G'(1)^2 \over p}   + B_1(y) G'(1)  \right) \left( {-\log( \log x)  \over \log p} - y \right) 
 \right]\, .
 \end{split}
\end{equation}
Now from (\ref{Ustar})   one has for large  negative $k$
\begin{equation} U  \left( e^{p^{k-y}} \right) =  p^{k-y} \left( - {G'(1) \over p } (k-y)   + B_1(y)    \right)
\, ,
\end{equation}
so that one can take the $n_0 \to \infty$ limit in the last expression of $V_3(x)$ to  get
\begin{multline}
V_3(x)= 
     -{ \log x \, \log( \log x)  \over  p \log p}     
  B_1(y) G'(1)   
 \\ 
+{G'(1)^2 \over 2 p^2 }  \ { \log x  \,\log( \log x)  \over \log p}   \left(   {\log( \log x)  \over \log p}  - 1\right) +             C_3(y) \log x 
\, ,
\label{V3-result}
\end{multline}
where the periodic function $C_3(y)$ is defined by 
\begin{multline}
             C_3(y) 
=   { 1  \over p}
\left[ 
-y B_1(y) G'(1)    - y (y+1) {G'(1)^2 \over 2  p }   
+\sum_{k \ge 0} p^{-k  + y}
 e^{p^{k-y}}  G' \left( e^{p^{k-y}} \right)  U  \left( e^{p^{k-y}} \right) \right.  \\
   \left.
+ \sum_{k =-\infty}^{-1} \left\{ p^{-k  + y}
  e^{p^{k-y}}  G' \left( e^{p^{k-y}} \right)  U  \left( e^{p^{k-y}} \right)    
   +{G'(1)^2 \over p } (k-y)   - B_1(y) G'(1)  \right\}  
\right]\, .
\label{C3-result}
\end{multline}
{ Here again one can check  directly in(\ref{C3-result})
 that $C_3 (y+1)=C_3(y)$.}

\bigskip

\section{Expected expression of $ \psi_\infty(\cdot)$}
\label{sec:expect}
Since here  $G(1)=0$ and $x_*=1$, the expected critical behavior  (\ref{crit-3}) of $ \psi_\infty(\cdot)$ should be of the form
\begin{eqnarray}
\psi_\infty(x) \simeq 
\left(\log x  \right)^\alpha B\left(- {\log  \left(\log x  \right) \over \log F'(1)} \right)  
 + O \left( \left(\log x  \right)^{\alpha +1} \right)  \, .
 \label{crit-4} 
\end{eqnarray}
For $\epsilon$ small one has 
\begin{equation} F'(1) = p + \epsilon f_1\, , \ \ \ 
 \alpha = 1 + \epsilon a_1 + \epsilon^2  a_2 + \ldots \ \text{ and } \ 
 B(y)=  1 + \epsilon B_1(y) + \epsilon^2  B_2(y) + \ldots\end{equation}
with
\begin{equation}f_1=  G'(1)  \, , \ \ \ 
a_1=  -  ~ {G'(1) \over p ~ \log p } 
\ \text{ and } \ 
a_2={  G'(1)^2  (2 + \log p ) \over 2  (p \log p)^2}\, , \end{equation}
and one gets from  (\ref{crit-4})
\begin{multline}
\psi_\infty(x)\, =\,  \log x + \left[  B_1(y) \log x + a_1 \log x  \, \log(\log x)) \right] \epsilon 
\\
  +   \left[    B_2(y) \log x 
 + a_2 \log(x) \,  \log(\log(x)) + a_1 B_1(y) \log x \,  \log(\log(x))
\right. \\  
\label{expected}
    \left.  + {a_1^2 \over 2} \log x \,  [\log(\log x)]^2 
 + f_1 {\log x \,  \log (\log x) \over p (\log p)^2}   B_1'(y) 
 \right] \epsilon^2  + \ldots  
 \end{multline} 
where
\begin{equation}
y\, =\, - {\log(\log x) \over \log p} \, .
\end{equation}

\medskip 

Comparing with (\ref{psi-inf})  and (\ref{expected})  with $U(x)$, $V_1(x)$, $V_2(x)$ and 
$V_3(x)$  replaced by their estimates \eqref{Ustar}, \eqref{V1-result}, \eqref{V2-result} and \eqref{V3-result}
one then finds 
\begin{eqnarray} 
B_2(y)= 
-{C_1(y) \over 2}+ 
B_1(y) C_2(y)
-{            C_3(y) }\, ,
\end{eqnarray} 
which leads to (\ref{B2-def}) once the periodic functions $C_1, C_2, C_3$ have been replaced by the expressions \eqref{C1-result},\eqref{C2-result} and \eqref{C3-result}.

\section{Large $p$ limit 
}
\label{sec:largep}
We consider now the 
the map \eqref{large-p2}: recall that $q\in(0,1)$ is fixed, as well as the polynomial $P(\cdot)$. 
When   $p$  is large $x_*=1$, $F'(1)=pq+(1-q)P'(1)\sim pq$ and   one has
(compare with \eqref{alpha})
\begin{equation}
\alpha\, =\, {\log p \over \log F'(1)}   \, \simeq\,  1 - {\log q \over \log p} \, .
\label{alphap}
\end{equation}
For  $x$ close to $1$, let us define $z$ by
\begin{equation}
1 + \left(F'(1)\right)^{n(x)} \  (x-1) \, =\,  1 + {z \over p}\, ,
\label{z-def}
\end{equation}
where $n(x)$ is the integer
such that
\begin{equation} 
p^{-\eta}  \, \le\,    z  \, < \,   F'(1)p^{-\eta} \,\stackrel{ p\to \infty}\sim \, q \, p^{1-\eta}\, ,
\end{equation}
and  $\eta\in (0,1)$ is some fixed number  (for example $\eta=1/3$).

The main idea is that for $x$ close to 1, the map \eqref{large-p2} can be replaced by
$F(x) \simeq 1 + F'(1) (x-1)$
for the first $n(x)$ iterations, and by
$F(x) \simeq q x^p$ beyond the $n(x)+2^{\text{nd}}$ iteration. Only the $n(x)+1^{\text{st}}$ iterate has 
to be calculated with the full expression \eqref{large-p2}. 

Therefore  
recalling that $P(1)=1$ we have
\begin{equation}
F^{n(x) + m +1} (x)  \simeq \Big( q \, e^{z}  + 1-q\Big)^{p^m} \, q^{{p^m-1 \over p-1}}\, ,
\end{equation}
and 
\begin{equation}
{\log F^{n(x) + m +1} (x) \over p^{n(x) + 1 + m}}\, =\, 
 {1 \over p} p^{-n(x)} \left[ \log \left(q \,  e^z + 1-q\right) 
 + \left( {1 \over p-1}  - {1 \over p^m (p-1)} \right) \log q \right]\, ,
 \end{equation}
and by taking the $m \to \infty$ limit and using $(F'(1))^\alpha=p$ we obtain
\begin{equation}
\psi_\infty(x)  \simeq  {(x-1)^\alpha  \over p}   \left({p \over z} \right)^\alpha \left[ \log
\left(q  \, e^z + 1-q \right) + {\log q \over p-1} \right]\, .
\end{equation}
Taking   the large $p$ limit
we therefore obtain
\begin{equation}
 \psi_\infty(x) \simeq (x-1)^\alpha  \   {\log
\left(q \,e^z + 1-q \right)  \over q \ z^\alpha}\, .  
\label{psi-p}
\end{equation}
One can then write
\begin{equation} - {\log( x-1) \over \log F'(1) } = n(x) + y \ \  \ \ \ {\rm with} \ \ \ \  \eta < y < \eta+1\, ,
\end{equation}
that is 
\begin{equation}
n(x) \, =\,   \left\lfloor - {\log( x-1) \over \log F'(1) } - \eta \right\rfloor\, ,
\end{equation}
where $\lfloor a \rfloor:= \max\{n\in \mathbb{Z}:\, n\le a\}$,
or equivalently (see \eqref{z-def}) 
\begin{equation}  
{z \over p} \, =\,  \left( F'(1)\right)^{-y}\, ,
\end{equation}
and note that  in the  large $p$ limit  $z\nearrow \infty$ (respectively $z \searrow 0$)
for $y<1$ (respectively $y>1$). 
Then one finds, for fixed $y \neq 1$
\begin{equation}
 \psi_\infty(x) \simeq \begin{cases} (x-1)^\alpha      q^{-y}  &  \ \ \ {\rm  for}   \ \  \eta\le  y < 1 \, ,
\\     (x-1)^\alpha q^{1-y} &   \ \ \ {\rm  for}  \ \  1<y < 1+\eta \, ,
\end{cases}
\end{equation}
as claimed in (\ref{lp}). 

At $y=1$ the amplitude becomes discontinuous.
One can resolve this discontinuity by analyzing, in the large $p$ limit, the range
$y-1 \sim (\log p)^{-1}$
and one gets (\ref{resolve}) from \eqref{alphap} and \eqref{psi-p}.

\section*{Acknowledgements}
The authors acknowledge the support of ANR, grant SHEPI.

\end{document}